\documentclass[12pt]{article}
\usepackage{epsf}
\title{{\bf Exotic baryon states in  QCD sum rule}}

\author{A.G.Oganesian\\
Institute of Theoretical and Experimental Physics,\\
B.Cheremushkinskaya 25, 117218 Moscow,Russia}
\date{}

\begin{document}
\pagestyle{empty}

\maketitle

\newcommand{\be}{\begin{equation}}
\newcommand{\ee}{\end{equation}}

\def\la{\mathrel{\mathpalette\fun <}}
\def\ga{\mathrel{\mathpalette\fun >}}
\def\fun#1#2{\lower3.6pt\vbox{\baselineskip0pt\lineskip.9pt
\ialign{$\mathsurround=0pt#1\hfil##\hfil$\crcr#2\crcr\sim\crcr}}}

\vspace{1cm}
\begin{abstract}
 It is shown, that the small decay width of $\Theta^+ = uudd\bar{s}$
baryon is suppressed by chirality violation. It is shown that
$\Theta^+$ decay width $\Gamma$ is proportional to $\alpha^2_s
\langle 0 \vert \bar{q} q \vert 0 \rangle^2$, for any pentaquark
current without derivatives.
\end{abstract}


\vspace{1cm}

\normalsize

In this talk we will discuss the the narrow exotic baryon
resonance $\Theta^+$ with quark content $\Theta^+ = uudd\bar{s}$
and mass 1.54 GeV. This resonance had been discovered last year by
two groups \cite{n1,n2}. Later, the existence of this resonance
was confirmed by many other groups, although some searches for it
were unsuccessful. (see \cite{n3} for the review). $\Theta^+$
baryon was predicted in 1997 by D.Diakonov, V.Petrov and
M.Polyakov \cite{n4} in the Chiral Soliton Model as a member of
antidecouplet with hypercharge $Y = 2$. The recent theoretical
reviews are given in \cite{n5,n6}. $\Theta^+$ was observed as a
resonance in the systems $nK^+$ and $pK^0$. No enhancement was
found in $pK^+$ mass distributions, what indicates on isospin
$T=0$ of $\Theta^+$ in accord with theoretical predictions
\cite{n4}.

My talk is based mainly on our paper \cite{nm} and I try to
explain some point of our paper more detailed. One of the most
interesting features of $\Theta^+$ is the very narrow width.
Experimentally, only an upper limit was found, the stringer bound
was presented in \cite{n2}: $\Gamma < 9 MeV$. The phase analysis
of $KN$ scattering results in the even stronger limit on $\Gamma$
\cite{n7}, $\Gamma < 1 MeV$. A close to the latter limitation was
found in \cite{n8} from the analysis of $Kd \to ppK$ reaction and
in \cite{n9} from $K+Xe$ collisions data \cite{n2}. The Chiral
Quark Soliton Model gives the estimation \cite{n4}: $\Gamma_{CQSM}
\la 15 MeV$ (R.E.Jaffe \cite{n10} claims that this estimation has
a numerical error and in fact $\Gamma_{CQSM}\la 30 MeV$ -- see,
however, \cite{n11}). In any way, such extremely narrow width of
$\Theta^+$ (less than $1 MeV$) seems to be very interesting
theoretical problem. My talk will be organized in the following
way: in the sect.1 I  suggest the qualitative explanation of the
narrow width of pentaquark and show that it is strongly
parametrically suppressed. It will be shown, that the conclusion
does not depend of the choice of the pentaquark current (without
derivatives). In sect. 2 we will discuss the possible pentaquark
currents and consider two-point correlation function.

\bigskip

{\bf \large Part 1. }

In this section  we will estimate the pentaquark width. Let us
consider 3-point correlator

\be
 \Pi_{\mu}=\int e^{i(p_1x-qy)} \langle 0\mid
\eta_{\theta}(x)j^5_{\mu}(y) \eta_n(0)\mid 0 \rangle \ee

where $\eta_n(x)$ is the neutron quark current \cite{n12},
($\eta_n =\varepsilon^{abc} (d^a C \gamma_{\mu} d^b) \gamma_5
\gamma_{\mu} u^c $),

$ \langle 0\mid \eta_n\mid n\rangle =\lambda_n v_n$, ($v_n$ and
$\lambda_n$ are nucleon spinor and nucleon transition constant
into nucleon current $eta_n$ ), $\eta_{\theta}$ is arbitrary
pentaquark current (not only $\Theta^+$, but with other isospin
also)
 $ \langle 0\mid \eta_{\theta}\mid \theta^+
\rangle=\lambda_{\theta} v_{\theta}$ and $j_{\mu 5} = \bar{s}
\gamma_{\mu}\gamma_5 u$ is the strange axial current.

As an example of $\eta_{\theta}$ one can use the following one:

\be
\eta_{\Theta}(x) = [\varepsilon^{abc} (d^a C \sigma_{\mu \nu} d^b)
\gamma_{\nu} u^c \cdot \bar{s} \gamma_{\mu} \gamma_5 u - (u
\leftrightarrow d)]/\sqrt{2}, \ee though all results in this
section are the same for any pentaquark current (without
derivatives). (Note, that this current have isospin 1, and I glad
to thank M. Nielsen, who pay my attention to this, but for any
current with isospin 0 and without derivatives  results will be
just the same).

As usual in QCD sum rule the physical representation of correlator
(1) can be saturated by lower resonances (both in $\eta_{\Theta}$
and nucleon channel)
\be
 \Pi^{Phys}_{\mu}=\langle 0\mid \eta_{\theta}\mid \theta^+ \rangle
\langle \theta^+ \mid j_{\mu}\mid n \rangle \langle n\mid \eta_n
\mid 0 \rangle \frac{1}{p^2_1 -m^2_{\theta}}\frac{1}{p^2_2-m^2}+
cont. \ee where $p_2=p_1-q$ is nucleon momentum, $m$ and
$m_{\theta}$ are nucleon and pentaquark masses.

Obviously,
\be
\langle \theta^+ \mid j_{\mu}\mid n \rangle =g^A_{\theta n}
\bar{v}_{\theta} \Biggl (g^{\mu\nu}
-\frac{q^{\mu}q^{\nu}}{q^2}\Biggr ) \gamma^v\gamma_5 v_n \ee where
axial transition constant $g^A_{\theta n}$ is just we are
interesting in (the width is proportional to the square of this
value). Such a method for calculation the width in QCD sum rules
is not new, see, e.g. \cite{n13}. Substituting all these in eq (3)
one can easily see, that 3-point correlator (1) is proportional to
$g^A_{\theta n}$.

Let us neglect quark masses and perform the chiral transformation
in (1) $q \to \gamma_5 q$. It is evident, that $\eta_n$ and
$j_{\mu 5}$ are even under such transformation, while
$\eta_{\Theta}$ is odd. Therefore, the correlator (1) vanishes in
the limit of chiral symmetry. It is easy to see, that this
statement is valid for any form of pentaquark and nucleon quark
currents (spinless and with no derivatives). In the real world the
chiral symmetry is spontaneously broken. The lowest dimension
operator, corresponding to violation of chiral symmetry is
$\bar{q}q$. So, the correlator (1) is proportional to quark
condensate $\langle 0 \vert \bar{q} q \vert 0 \rangle$. Just the
same result one can found from direct calculation of invariant
amplitude (at convenient kinematical structure, for example $\hat
{p} p^{\mu}$). So we come to conclusion, that $\Theta^+$ width is
suppressed by chirality violation for any pentaquark current
without derivatives(i.e. axial transition constant should be
proportional to quark condensate).

The second source of the suppression also does not depend on the
form of the pentaquark current. Let us again consider correlator
(1). One can easily note, that unit operator contribution to this
correlator (bare diagram) is expressed in the terms of the
following integrals
\be
\int e^{i(p_1x-qy)} \frac{d^4 x d^4 y}{((x-y)^2)^n (x^2)^m} \equiv
\int \frac{e^{ip_1x}}{(x^2)^m} \frac{e^{-iq t}}{(t^2)^{n}} d^4
xd^4t \ee

It is clear that such integrals have imaginary part on $p_2^2$ and
$q^2$ - the momentum of nucleon and axial current - but there is
no imaginary part on $p_1^2$ - the momentum of pentaquark. So we
come to the conclusion that bare diagram  correspond to the case,
when there is no $\Theta^+$ resonance in the pentaquark current
channel (this correspond to background of this decay). (Note, that
this conclusion don't depend on the fact that one of the quark
propagators should be replaced by condensate, as we discuss
before). The imaginary part on $p_1^2$ (i.e. $\Theta^+$
resonance)appears only if one take into account hard gluon
exchange. So we come to conclusion, that if $\Theta^+$ is a
genuine 5-quark state (not, say, the $NK$ bound state), then in
(2) the hard gluon exchange is necessary, what leads to additional
factor of $\alpha_s$. We come to the conclusion, that
$\Gamma_{\Theta} \sim \alpha^2_s \langle 0 \vert \bar{q} q \vert 0
\rangle^2$, i.e., $\Gamma_{\Theta}$ is strongly suppressed. This
conclusion takes place for any genuine 5-quark states -- the
states formed from 5 current quarks at small separation, but not
for potentially bounded $NK$-resonances, corresponding to large
relative distances. There are no such suppression for the latters.
I want to repeat once more, that this conclusion don't depend on
the choice of current (without derivatives).

(I would like to add that recently (after this talk was given) in
 the paper of D.Melikhov and B.Stech \cite{n21}, the pentaquark in
 the Chiral symmetry limit was investigated and authors note that
 they results agree with our conclusion about pentaquark width suppression due
 to chirality violation).

Author thanks K.Goeke and to M.Polyakov for useful discussions and
for their kind hospitality in Bochum university and  M.Nielsen for
significant note.

 This work was supported in part by INTAS grant 2000-58 and by RFBR grant
03-02-16209.

\end{document}